\documentclass[showpacs,amsmath,amssymb,twocolumn,pra,longbibliography,superscriptaddress]{revtex4-2}
\usepackage{siunitx}
\usepackage{amssymb}
\usepackage[dvips]{graphicx}
\usepackage{enumerate}
\usepackage{epsfig} 
\usepackage{subfigure}
\usepackage{xcolor}
\usepackage[T1]{fontenc}
\usepackage{fullpage}
\usepackage{amsthm,amsfonts,amssymb,amscd,mathrsfs,xspace,framed}
\usepackage{mathrsfs,amsmath}
\usepackage{color}
\usepackage{setspace}
\usepackage{url}
\usepackage{wrapfig}
\usepackage{tikz}
\usepackage{enumitem}
\usepackage{bm}
\usepackage{txfonts}
\usepackage{array}
\usepackage{color}
\usepackage{braket}
\usepackage{natbib}
\usepackage{changes}
\usepackage{soul}
\usepackage[colorlinks=true,citecolor=blue,linkcolor=magenta]{hyperref}
\usepackage{cleveref}
\usepackage{graphicx}

\def\be{\begin{equation}}
\def\ee{\end{equation}}
\def\ba{\begin{eqnarray}}
\def\ea{\end{eqnarray}}

\newcommand{\EPFL}{Swiss Federal Institute of Technology Lausanne (EPFL), CH-1015 Lausanne, Switzerland}
\newcommand{\CfQS}{Center for Quantum Science and Engineering, EPFL, Lausanne, Switzerland}
\newcommand{\SEAS}{John A. Paulson School of Engineering and Applied Sciences, Harvard University, Cambridge, MA, USA.}
\newcommand{\CTH}{Department of Microtechnology and Nanoscience (MC2),
	Chalmers University of Technology, SE-412 96 G\"oteborg, Sweden}

\begin{document}

\title{Motional sideband asymmetry of a solid-state mechanical resonator at room temperature}

\author{Yi Xia}
\email{yi.xia@epfl.ch}
\thanks{These authors contributed equally.}
\affiliation{\EPFL}
\affiliation{\CfQS}

\author{Guanhao Huang}
\thanks{These authors contributed equally.}
\affiliation{\EPFL}
\affiliation{\CfQS}
\affiliation{\SEAS}
\author{Alberto Beccari}
\affiliation{\EPFL}
\affiliation{\CfQS}

\author{Alessio Zicoschi}
\affiliation{\EPFL}
\affiliation{\CfQS}
\author{Amirali Arabmoheghi}
\affiliation{\EPFL}
\affiliation{\CfQS}
\author{Nils J. Engelsen}
\affiliation{\EPFL}
\affiliation{\CfQS}
\affiliation{\CTH}

\author{Tobias J. Kippenberg}
\email{tobias.kippenberg@epfl.ch}
\affiliation{\EPFL}
\affiliation{\CfQS}

\begin{abstract}
The motional sideband asymmetry of a mechanical oscillator interacting with a laser field can be observed when approaching the quantum ground state, where the zero-point energy of the mechanical oscillator becomes a sizable contribution to its motion. In the context of quantum optomechanics,  it allows, in principle, calibration-free inference of the thermal equilibrium of a macroscopic mechanical resonator with its optical bath. At room temperature, this phenomenon has been observed in pioneering experiments using levitated nanoparticles.  Measuring this effect with solid-state mechanical resonators has been compounded by thermal intermodulation noise, mirror frequency noise and low quantum cooperativity. Here, we sideband-cool a membrane-in-the-middle system close to the quantum ground state from room temperature, and observe motional sideband asymmetry in a dual-homodyne measurement.  Sideband thermometry yields a minimum phonon occupancy of $\Bar{n}_\mathrm{eff}=9.5$.  Our work provides insights into nonlinear optomechanical dynamics at room temperature and facilitates accessible optomechanical quantum technologies without the need for complex feedback control and cryogenic cooling.
\end{abstract}

\maketitle

A genuine non-classical feature of a quantum harmonic oscillator with angular frequency $\Omega$ is its zero-point energy $\hbar\Omega/2$ even at vanishing temperature. In cavity optomechanics, this is revealed by sideband thermometry, as the nonidentical spectral heights of the optical sidebands at positive and negative frequencies $\pm\Omega$~\cite{aspelmeyer2014cavity}, with their power ratio $(\bar{n}+1)/\bar{n}$, encodes the mean phonon occupancy $\bar{n}$. For frequency-resolved detection of Stokes and anti-Stokes photons ~\cite{Galinskiy20}, sideband asymmetry directly reflects the ability of the mechanical resonator to emit or absorb energy ~\cite{RevModPhys.82.1155}, as in molecular systems~\cite{kip1990determination,cui1998noncontact}. In the framework of linear measurements, the different spectral heights can be interpreted as the measurement-induced imprecision-backaction correlation from the optical field ~\cite{Weinstein14,khalili2012quantum}. Nonetheless, this ratio significantly deviates from unity when the mechanical oscillator approaches the quantum ground state. However, the oscillator's zero-point fluctuations is typically obscured by the excess thermal energy ($k_\mathrm{B}T\gg \hbar\Omega$) from coupling with the environmental bath at temperature $T$, making it challenging to enter the quantum regime, in particular for low frequency mechanical oscillators.

Over the last decades, laser cooling ~\cite{schliesser2008resolved,wilson2015measurement} and cryogenic refrigeration have been combined to prepare the motional quantum ground state of mechanical oscillators~\cite{chan2011laser,rossi2018measurementbased,tebbenjohanns2021quantum} which is a key prerequisite for many quantum protocols, such as mechanical squeezing~\cite{szorkovszky2011mechanical}, entanglement~\cite{riedinger2018remote}, and non-classical state preparation~\cite{Hong17}. Developing optomechanical systems that can reach the quantum regime at room temperature helps reduce the experimental complexity and is beneficial for the widespread adoption of optomechanical quantum technologies, such as hybrid quantum systems ~\cite{moller2017quantum} and quantum sensing ~\cite{mason2019continuous,xia2023entanglement}. To date, only a few optomechanical systems have reached the quantum regime at room temperature: levitated nanoparticles~\cite{Delic20,Tebbenjohanns20,magrini2021realtime}, optical spring-stiffened micromirrors~\cite{cripe2019measurement} and solid-state nanomechanical resonators~\cite{Huang2024}. Yet, ground-state cooling and measurement of sideband asymmetry have only been demonstrated with levitated nanoparticles~\cite{Tebbenjohanns20,Delic20,magrini2021realtime}. For solid-state mechanical resonators~\cite{saarinen2023laser}, approaching the quantum ground state has remained elusive, as the low quantum cooperativity limits the coupling to the cold optical bath~\cite{guo2019feedback,saarinen2023laser}, while excess classical noise -- such as cavity frequency noise --  thermalizes the oscillators~\cite{Pluchar23}. Other compounding factors include thermal instabilities~\cite{PhysRevLett.101.133903} and intermodulation noise~\cite{fedorov2020thermal}.

In this letter, we demonstrate optomechanical sideband asymmetry with a sideband-cooled membrane oscillator close to the quantum ground state at room temperature. We cool the mechanical oscillator with a strong pump laser at the `magic' detuning to minimize the thermal intermodulation noise (TIN), which originates from the nonlinear optomechanical transduction response of a narrow-linewidth optical resonance ~\cite{fedorov2020thermal}. We observe significant sideband asymmetry utilizing a novel dual-homodyne scheme. By using a weak probe laser with a large-linewidth optical resonance, we avoid nonlinear noise contamination in the sideband asymmetry measurement. Our result elucidates the influence of nonlinear cavity transduction noise on optomechanical sideband asymmetry measurements, and provides a path to more closely approach the quantum ground state of solid-state mechanical oscillators at room temperature. 


\begin{figure}
    \centering
    \includegraphics[width=0.5\textwidth]{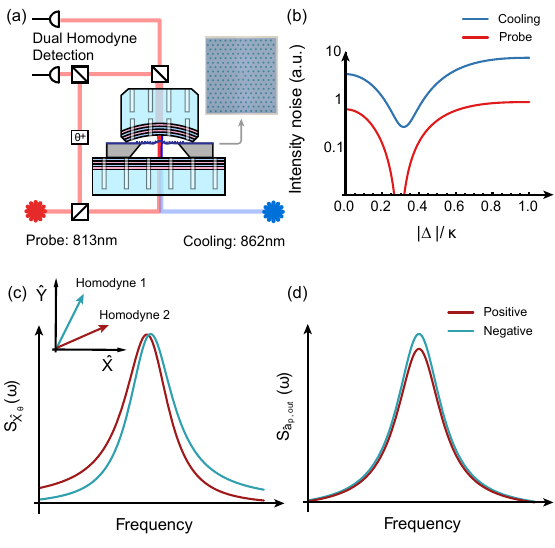}
    \caption{\textbf{Sideband asymmetry measurement from a dual-homodyne detection}. (a) Experiment setup. The defect mode of a density-modulated membrane is sideband-cooled by a laser at \SI{862}{\nano\meter} and probed with a laser at \SI{813}{\nano\meter}.  (b) Intracavity intensity noise from TIN as a function of cavity detuning. Intensity noise from the TIN of the cooling laser (blue curve) cannot be canceled completely at the magic detuning ($\Delta=-\kappa/2\sqrt{3}$). (c) Individual power spectral density (PSD) of dual homodyne detectors with different measurement angles. (d) Reconstructed power spectral density of output probe field $\bar{S}_{\hat a_\mathrm{p, out}\hat a_\mathrm{p, out}}(\omega) $ from dual-homodyne measurement. }
    \label{fig:1}
\end{figure}
The essentials of our experimental setup are shown in Fig.~\ref{fig:1} (a) and are described in more detail in ~\cite{Huang2024}. The mechanical resonator is a density-modulated phononic $\mathrm{Si_3N_4}$ membrane~\cite{Hoj24} patterned with amorphous silicon (aSi) pillars to create an acoustic bandgap near \SI{1.15}{MHz}. At the center of the membrane, a defect supports a soft-clamped, out-of-plane flexural mode with mechanical frequency $\Omega_m/2\pi = \SI{1.167}{MHz}$ and quality factor $Q = 1.8\times10^8$. The membrane is fixed in an optical Fabry–Pérot cavity consisting of two phononic-crystal-patterned mirrors which reduce the cavity frequency noise around the mechanical frequency. 

A strong cooling laser with input power around 10 mW  at \SI{862}{nm} is red-detuned from the cavity resonance by $\Delta=-\kappa/2\sqrt{3}$ (magic detuning; where $\kappa/2\pi= \SI{14}{\mega\hertz}$ is the cavity linewidth at 860nm) to suppress the TIN intracavity photon number fluctuations~\cite{Pluchar23}. In the limit of linearized optomechanical coupling~\cite{aspelmeyer2014cavity}, the mechanical vibration dispersively modulates the optical cavity frequency at $\pm\Omega_m$, and the detuning from the cavity resonance enhances the anti-Stokes scattered light at rate $A^- = g^2\kappa|\chi_\mathrm{cav}(\Omega_m)|^2$ while suppressing the Stokes scattered light at rate $A^+=g^2\kappa|\chi_\mathrm{cav}(-\Omega_m)|^2$ with the cavity susceptibility $\chi_\mathrm{cav}(\omega)^{-1} = \kappa/2-i(\omega+\Delta)$ and the enhanced optomechanical coupling rate $g$. The net effect $\Gamma_\mathrm{opt}=A^- - A^+$ of the two processes removes the mechanical thermal quanta when the pump light leaves the cavity until an equilibrium value $\bar{n}_\mathrm{eff} = \frac{A^++n_\mathrm{th}\Gamma_\mathrm{m}}{\Gamma_\mathrm{opt}+\Gamma_\mathrm{m}}$ is reached with both thermal environment $\bar{n}_\mathrm{th}$ and the cold optical bath with quantum-limited laser noise.  

In a highly multimode optomechanical system with cavity frequency excursions that are large compared to the cavity linewidth, the linearized regime of optomechanics breaks down~\cite{fedorov2020thermal}. The cavity frequency fluctuations, including those due to mechanical modes outside the acoustic bandgap, can go through nonlinear frequency mixing and create classical noise (TIN)  much larger than the laser shot noise even within the acoustic bandgap. In the fast cavity limit $\kappa \gg \Omega_m$, the quadratic transduction-induced radiation pressure fluctuation can be canceled exactly at the magic detuning~\cite{fedorov2020thermal}, and higher-order transduction noise is negligible. For sideband cooling, where a narrow cavity linewidth is preferred, the residual quadratic and higher-order TIN~\cite{huang2024room} (see Sec. A.1. of SI) remain even at the magic detuning (shown in Fig.~\ref{fig:1} (b)). This excess nonlinear radiation pressure noise can drive the mechanical mode differently from any linear noise source. The noise-driven motion further interferes with the nonlinear noise at the detector (see Sec. A.1. and B.3. of SI), making retrieval of the phonon occupation by spectral fitting challenging, and preventing sideband cooling from approaching the quantum backaction limit $\bar{n}_{\mathrm{min}}=[(\Omega_m+\Delta)^2+(\kappa)^2]/(-4\Delta\Omega_m)$.


\begin{figure*}
    \centering
    \includegraphics[width=1\textwidth]{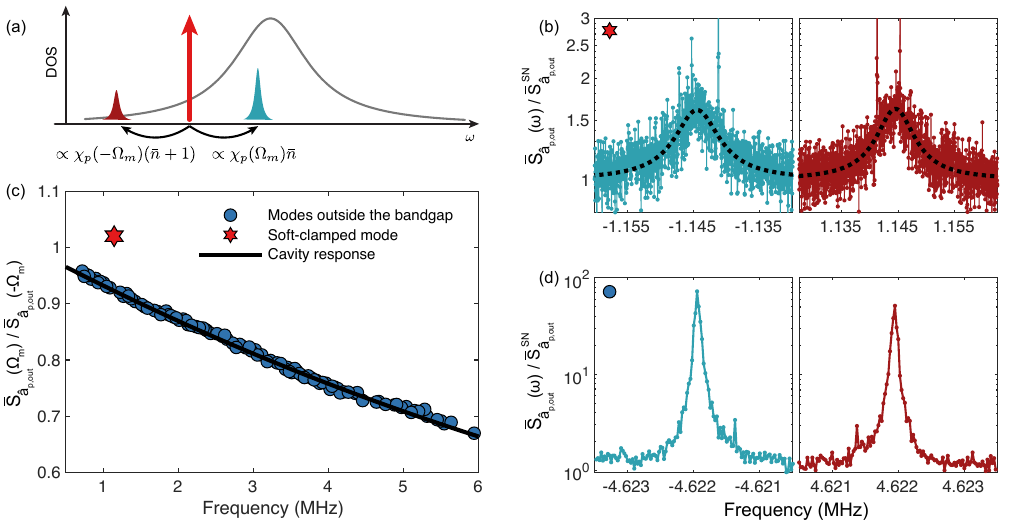}
    \caption{\textbf{Observation of motional sideband asymmetry at room temperature from a solid state density modulated membrane defect mode}. (a) Stokes and anti-Stokes scattering rates depend on both the cavity response and the mean phonon occupation. (b, d) Power spectral density  $\bar{S}_{\hat a_\mathrm{p, out}\hat a_\mathrm{p, out}}(\omega) $ obtained from dual homodyne detection, normalized to shot noise (SN), showing the spectral power of the defect mode (b) and a mechanical mode outside the band gap (d) at positive and negative frequencies. The spurious narrow peaks atop the mechanical spectrum originate from electrical signals and are excluded from the fitting (dashed lines). (c) Sideband asymmetry of soft-clamped defect mechanical mode (red hexagram) and  100 mechanical modes outside the bandgap (blue circles).  The asymmetry of defect mode strongly deviates from the cavity susceptibility-induced asymmetry (black line). }
    \label{fig:2}
\end{figure*}

We calibrate the phonon occupation by measuring the sideband asymmetry with a weak probe laser at \SI{813}{\nano\meter} in the fast cavity limit. This weak probe is red-detuned from the cavity resonance at the magic detuning  $\Delta_\mathrm{p}=-\kappa_\mathrm{p}/2\sqrt{3}$ ($\kappa_\mathrm{p}/2\pi=49.4$ MHz is the cavity linewidth at 813 nm). This configuration is chosen such that the residual quadratic and higher-order TIN can be efficiently eliminated from the probe beam, making phonon occupancy measurements more reliable.  The TIN in the measurement of the probe beam can be efficiently canceled at arbitrary optical quadrature angles in a homodyne measurement by adjusting the appropriate local oscillator amplitude and phase in a single-port homodyne scheme~\cite{Huang2024}. However, TIN cancellation in a heterodyne measurement~\cite{huang2024room}, typically utilized for probing sideband asymmetry, has been challenging to implement. 

We use a dual-homodyne scheme (shown in Fig.~\ref{fig:1} (c,d)) to measure the optomechanical sideband asymmetry without thermal intermodulation noise contamination, via appropriate measurement configurations to cancel the nonlinear noise in each detection port. In Sec. I.C. of SI, we show that the dual-homodyne scheme effectively realizes a heterodyne measurement of the optical field, yet allows for sufficient degrees of freedom to completely cancel the TIN in detection. 

To implement the scheme, the probe light from the membrane-in-the-middle (MIM) setup is split evenly into two optical ports. At one detection port, the amplitude quadrature $X_{\theta_1=0}$ is measured with direct detection where the TIN on the amplitude quadrature is naturally canceled at the magic detuning. At the other detection port, the phase quadrature $X_{\theta_2=-\pi/3}$ is acquired with single-port homodyne detection, where the TIN on the phase quadrature is eliminated by using an appropriate LO amplitude\cite{fedorov2020thermal}. The photocurrent streams from the two detectors are digitized simultaneously and post-processed to reconstruct the symmetrized power spectral density (PSD) $ \bar{S}_{\hat a_\mathrm{p, out}\hat a_\mathrm{p, out}}(\omega) = \frac{1}{2}\left({S}_{\hat a_\mathrm{p, out}\hat a_\mathrm{p, out}}(\omega) + {S}_{\hat a_\mathrm{p, out}^\dag\hat a_\mathrm{p, out}^\dag}(-\omega)\right)$
of the output probe field $\hat a_\mathrm{p, out}$, equivalent to that of a heterodyne measurement of the field.

\begin{figure*}
    \centering
    \includegraphics[width=1\textwidth]{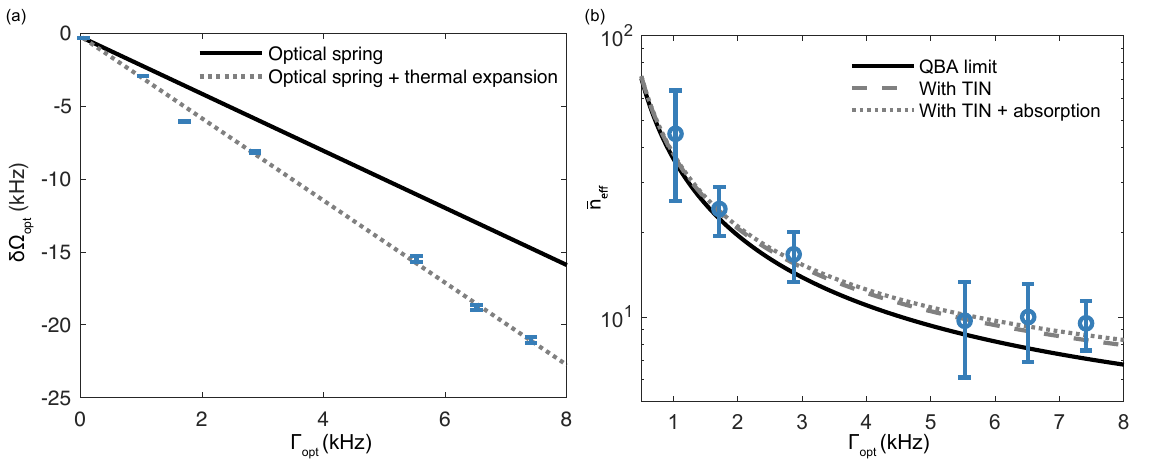}
    \caption{\textbf{Sideband cooling limit.} (a) Mechanical resonance frequency shifts as a function of optomechanical damping rate, due to optical spring shift and optical absorption. Raising the cooling power increases the optomechanical damping rate, leading to a reduction of the effective temperature of the defect mode.  (b) Phonon occupancy retrieved from sideband asymmetry at different optomechanical damping rates. The measured phonon occupation is above the quantum backaction limit due to the excess heating from residual TIN and absorption of  cooling beam. Error bars indicate the standard errors in spectral fittings.}
    \label{fig:3}
\end{figure*}

The reconstructed PSDs contain Lorentzian peaks at mechanical frequencies $\Omega_m$ on top of the vacuum noise of the probe light, whose positive and negative frequency components are modified by the cavity susceptibility (see Sec. A.3. of SI) 
\begin{gather}
    \bar{S}_{\hat a_\mathrm{p, out}\hat a_\mathrm{p, out}} (\omega) \nonumber \\
    = 2\eta g_\mathrm{p}^2\kappa_{\mathrm{p},1}|\chi_\mathrm{p}(-\omega)|^2(\bar{S}_{\hat Q\hat Q}(\omega)+ \mathrm{Im}\{\chi_\mathrm{m}'(\omega)\})+1/2,
\end{gather}
where $g_\mathrm{p}$ is the optomechanical coupling rate of the probe light, $\eta$ is the measurement efficiency, $\chi_\mathrm{p}^{-1} =\kappa_\mathrm{p}/2-i(\omega+\Delta_\mathrm{p})$ is the inverse probe cavity susceptibility, $\bar{S}_{\hat Q\hat Q}$ is the PSD of the dimensionless mechanical position operator $\hat Q$ (including effects from laser cooling and noise heating by the residual TIN at the cooling cavity resonance), and $\chi_\mathrm{m}'$ is the modified mechanical susceptibility due to optomechanical coupling with the cooling and the probe lasers. The $\mathrm{Im}\{\chi_\mathrm{m}'(\omega)\}$ term comes from the imprecision-backaction correlation of the probe field~\cite{Weinstein14,khalili2012quantum} which gives rise to the measured optical sideband asymmetry ratio $R(\Omega'_m) = r(\bar{n})s(\Omega'_m)$ at shifted positive and negative mechanical frequencies $\pm\Omega'_m=\pm(\Omega_m+\delta\Omega_{\mathrm{opt}})$ (shown in Fig.~\ref{fig:2} (a)), with the mechanical sideband ratio $r(\bar{n})=(\bar{n}+1)/\bar{n}$ in addition to a cavity susceptibility-induced asymmetry factor
\begin{gather}
    s(\Omega'_m) =|\chi_\mathrm{p}(-\Omega'_m,\Delta_\mathrm{p})|^2/|\chi_\mathrm{p}(\Omega'_m,\Delta_\mathrm{p})|^2 ,
\end{gather}
that results from a non-zero detuning of the probe laser. For non-soft-clamped mechanical modes outside the acoustic bandgap, the final phonon occupancy $\bar{n}_\mathrm{eff} \gg 1$. Therefore, the sideband ratio at their resonant frequency follows exactly the cavity susceptibility-induced asymmetry $s(\Omega'_m)$, making them excellent resources for system calibration. As evidenced in Fig.~\ref{fig:2} (c), the ratios between the positive and negative sideband peaks from 100 mechanical modes align excellently with the cavity response, reflecting the difference in the cavity density of states. Strikingly, for the soft-clamped defect mode with low phonon occupancy, the sideband asymmetry ratio $R(\Omega'_m)$ significantly deviates from the cavity response $s(\Omega'_m)$. As shown in Fig.~\ref{fig:2} (b),  we observe nearly equal positive and negative spectral densities, which are due to the additional asymmetry $r(\bar{n})$ introduced by the low thermal occupancy of the defect mode.
Thus, the phonon occupation $\bar{n}_\mathrm{eff}^{-1}=1-R/s$ can be retrieved with $s$ precisely calibrated by the non-soft-clamped modes. Here we find a minimum final phonon occupation $\bar{n}_\mathrm{eff}=9.5\pm1.9$. 

In the following, we characterize the sideband cooling limit of our system. To this end, we measure the sideband asymmetry as a function of cooling power and compare the inferred phonon occupation from the sideband asymmetry measurements with those expected at the quantum backaction limit. As the cooling power increases, the mechanical frequency shifts to lower frequencies and the linewidth broadens due to dynamical backaction~\cite{aspelmeyer2014cavity}. 
Fig.~\ref{fig:3} (a) shows mechanical frequency shifts $\delta\Omega_{\text{opt}}$ at different optomechanical damping rates $\Gamma_{\text{opt}}$ which are retrieved by fitting the measured PSDs of the probe light.  Apart from the optical spring shift, we observe an additional frequency shift proportional to the cooling power. We attribute this to optical absorption, where thermal expansion induces stress relaxation of the membrane~\cite{Land24}, as the maximum intra-cavity optical power is $\mathcal{O}$(1) Watt. Because of the increasing temperature of the environment, the soft-clamped mode reaches thermal equilibrium with a hotter thermal bath at higher optical power. We simulate the mechanical frequency shift per temperature change at room temperature to be around \SI{0.67}{kHz/K}, which suggests a $7\%$ increase of the initial thermal occupation at the maximum optomechanical damping rate of \SI{7.4}{\kilo\hertz}. The high thermal conductance of the density-modulated membrane reduces the photothermal effect in our system. The TIN-induced nonlinear radiation pressure noise from the cooling beam is found to be another decoherence source. By direct detection of the output beam from the pump cavity mode, we have direct access to the intracavity nonlinear radiation pressure noise, which is measured to be around $15\%$ of the laser shot noise (See Sec. A.1. of SI). 

In Fig.~\ref{fig:3} (b), we plot the phonon occupation retrieved from the sideband asymmetry measurement at different optomechanical damping rates and compare them with our theoretical model that includes excess thermal decoherence from TIN and optical absorption. At the highest powers employed in the experiment, the measured phonon occupation is observed to be slightly above the quantum backaction limit~\cite{Peterson2016} ($\sim 7$ phonons), consistent with our decoherence model. At even higher cooling powers, the mechanical frequency is down-shifted to the edge of the acoustic bandgap, limiting the maximum achievable damping rate.


In conclusion, we report quantum sideband asymmetry measurements at room temperature. Instead of photon counting~\cite{Galinskiy20} and heterodyne detection~\cite{Delic20,Tebbenjohanns20}, we implement a novel dual-homodyne detection to measure the sideband asymmetry without TIN contamination of the measurement record. The appearance of sideband asymmetry is a direct consequence of the low phonon occupation, which is 6 orders of magnitude lower than that of room temperature.  Combining real-time optimal state estimation~\cite{magrini2021realtime} and feedback control~\cite{rossi2018measurementbased} using cavity modes with larger linewidths,
our optomechanical system could be placed deep in the quantum ground state at room temperature, comparable to the levitated nanoparticles~\cite{Delic20,magrini2021realtime}, yet with an effective mass 6 orders of magnitude larger. Our work paves the way for quantum control of solid-state mechanical resonators without cryogenic cooling and facilitates the development of hybrid quantum systems and accessible optomechanical quantum technologies.

\section*{acknowledgements}

We thank Sergey A. Fedorov for setting up the initial experiment. This work was supported by funding from the Swiss National Science Foundation under grant agreement no. 216987 (Postdoctoral Fellowship), grant agreement no. 185870 (Ambizione) and grant agreement no. 204927 (Cavity Quantum Electro-optomechanics).  This work has also received funding from the European Research Council (ERC) under the EU H2020 research and innovation programme, grant agreement No. 835329 (ExCOM-cCEO) . All samples were fabricated in the Center of MicroNanoTechnology (CMi) at EPFL. 

\section*{Author contribution} 
Y.X., G.H. performed the experiments, and the data analysis with assistance from A.A. and A.Z.. G.H., Y.X. conceived the theoretical framework, built the experiment setup, with assistance from N.J.E. The membrane sample is designed and fabricated by A.B. The manuscript was written by Y.X., G.H. with input from all others.  T.J.K. and Y.X. supervised the project.

\section*{Data Availability Statement} 
The code and data used to produce the plots within this work will be released on the repository \texttt{Zenodo} upon publication of this preprint.

\begin{appendix}
\onecolumngrid

\section{Theoretical framework}
\subsection{Residual thermal intermodulation noise}

Thermal intermodulation noise (TIN) comes from the mixing of different Fourier components of cavity frequency fluctuations due to nonlinear cavity transduction. Starting from the classical Langevin equation of cavity field $a$:
\begin{equation}
    \dot{a}(t) = [i(\Delta_0+\Delta(t))-\kappa/2]a(t)+\sqrt{\kappa_{\mathrm{in}}}a_{\mathrm{in}}(t)
\end{equation}
where $a_{\mathrm{in}}$ is the input optical field and $\Delta_0$ is the mean cavity detuning. We also ignore other loss channels for simplicity. The solution is expressed in the Liouville-Neumann series of multiple orders:
\begin{align}
    a &= \bar{a}\delta(\omega)+ a_1(\omega)+a_2(\omega)+a_3(\omega)+...+a_n(\omega)+... \\
     a_n(\omega)&=\bar{a} \iint \frac{i \Delta\left(\omega-\omega_1\right)}{i(-\omega-\Delta_0)+\kappa / 2} \frac{i \Delta\left(\omega_1-\omega_2\right)}{i\left(-\omega_1-\Delta_0\right)+\kappa / 2} \cdots \frac{i \Delta\left(\omega_{n-1}\right)}{i\left(-\omega_{n-1}-\Delta_0\right)+\kappa / 2} \frac{d \omega_1}{2 \pi} \cdots \frac{d \omega_{n-1}}{2 \pi} = \Bar{a}\delta a_n.
\end{align}
with  $\bar{a} =\frac{\sqrt{\kappa_p}\bar{a}_p}{-i\Delta_0+\kappa / 2} $.

The intracavity field intensity can be written as:
\begin{align}
    I  = &(a^\dagger_0+a^\dagger_1+a^\dagger_2+...)(a_0+a_1+a_2+...) \\
    =  &|\bar{a}|^2(\delta(\omega)+\delta a^\dagger_1+\delta a_1+\delta a^\dagger_2+\delta a_2+\delta a^\dagger_1\delta a_1 +...\\
       = & I^{(0)}+I^{(1)}+I^{(2)}...
    \end{align}
In the fast cavity limit, by ignoring the frequency dependence of $\omega$ in the cavity response ($i(-\omega-\Delta)+\kappa/2 \rightarrow i(-\Delta)+\kappa/2$),  the intensity fluctuations of intracavity field, up to the second order TIN, is thus:
\begin{align}
    I(\omega) &= |\bar{a}|^2(\delta(\omega)+ \delta a_1(\omega)+\delta a^\dagger_1(\omega)+\int \delta a^\dagger_1(\omega-\omega')\delta a_1(\omega') d\omega' +\delta a_2(\omega)+\delta a^\dagger_2(\omega)) \\
    & = |\bar{a}|^2 (\delta(\omega)-\frac{2\Delta(\omega)}{\Delta_0^2+(\kappa/2)^2} +\frac{3\Delta_0^2-(\kappa/2)^2}{[\Delta_0^2+(\kappa/2)^2]^2} \int \Delta(\omega-\omega')\Delta(\omega')d\omega')
\end{align}
At detuning $\Delta_0 = \pm\kappa/(2\sqrt{3})$ (magic detuning), the second-order TIN is canceled. However, this is no longer valid in sideband cooling where $\kappa \sim \omega$. In addition, the third-order TIN starts to become significant as the denominator in Eq.(S3) decreases.

We can estimate the residual second-order TIN by including the cavity response:
\begin{align}
    I^{(2)}(\omega) &= |\bar{a}|^2 (\int \delta a^\dagger_1(\omega-\omega')\delta a_1(\omega') d\omega' +\delta a_2(\omega)+\delta a^\dagger_2(\omega)) \\
    & \propto |\bar{a}|^2 (\int  G(\omega-\omega',\omega') \Delta(\omega-\omega')\Delta(\omega')) d\omega'
\end{align}
where $G(\omega-\omega',\omega')=\frac{1}{i(-\omega+\omega'+\Delta)+\kappa/2}\frac{1}{i(-\omega'-\Delta)+\kappa/2}
    -\frac{1}{i(-\omega-\Delta)+\kappa/2}\frac{1}{i(-\omega'-\Delta)+\kappa/2}-\frac{1}{i(-\omega+\Delta)+\kappa/2}\frac{1}{i(\omega'+\Delta)+\kappa/2}$.  The intensity fluctuation power spectral density of the second order TIN is:
\begin{align}
    S^{(2)}_{II} (\omega) &= FT_\tau \braket{I^{(2)}(t+\tau)I^{(2)}(t)} \\
 &\propto FT_\tau  \int\int\int\int \braket{\Delta(t-t_1+\tau)\Delta(t-t_2+\tau) \Delta(t-t_3)\Delta(t-t_4)}G(t_1,t_2)G(t_3,t_4) dt_1 dt_2 dt_3 dt_4 \\
 &\propto \braket{I(t)}^2 + FT_\tau \int\int\int\int (\braket{\Delta(t-t_1+\tau)\Delta(t-t_3)}\braket{\Delta(t-t_2+\tau)\Delta(t-t_4)}\\&+\braket{\Delta(t-t_1+\tau)\Delta(t-t_4)}\braket{\Delta(t-t_2+\tau)\Delta(t-t_3)}) G(t_1,t_2)G(t_3,t_4) dt_1 dt_2dt_3 dt_4 \\
 & \propto \int S_{\Delta\Delta}(\omega-\omega')S_{\Delta\Delta}(\omega')[G(\omega-\omega',\omega')G(\omega'-\omega,-\omega')+G(\omega-\omega',\omega')G(-\omega',\omega'-\omega)] d\omega'
\end{align}

\noindent where $FT$ stands for Fourier transform.

Third-order TIN can be calculated in a similar way. The full expression is cumbersome and we write its contribution in the bad cavity limit: 
\begin{align}
I^{(3)} &= |\bar{a}|^2 (\delta a_1^\dagger\delta a_2+ \delta a_2^\dagger\delta a_1+\delta a_3+\delta a^\dagger_3) \\
& \propto |\bar{a}|^2\Re \{i|\chi_\mathrm{cav}|^2\chi_\mathrm{cav}-i\chi_\mathrm{cav}^3\} \delta\Delta^3(t)
\end{align}
with the cavity susceptibility: 
\begin{align}
     \chi^{-1}_{\mathrm{cav}}= -i(\Delta_0+\omega)+\kappa/2
\end{align}
At the magic detuning, the third-order TIN still retains a finite value and can be comparable to the residual second-order TIN in the good cavity limit.
\begin{figure*}
    \centering
    \includegraphics[width=1\textwidth]{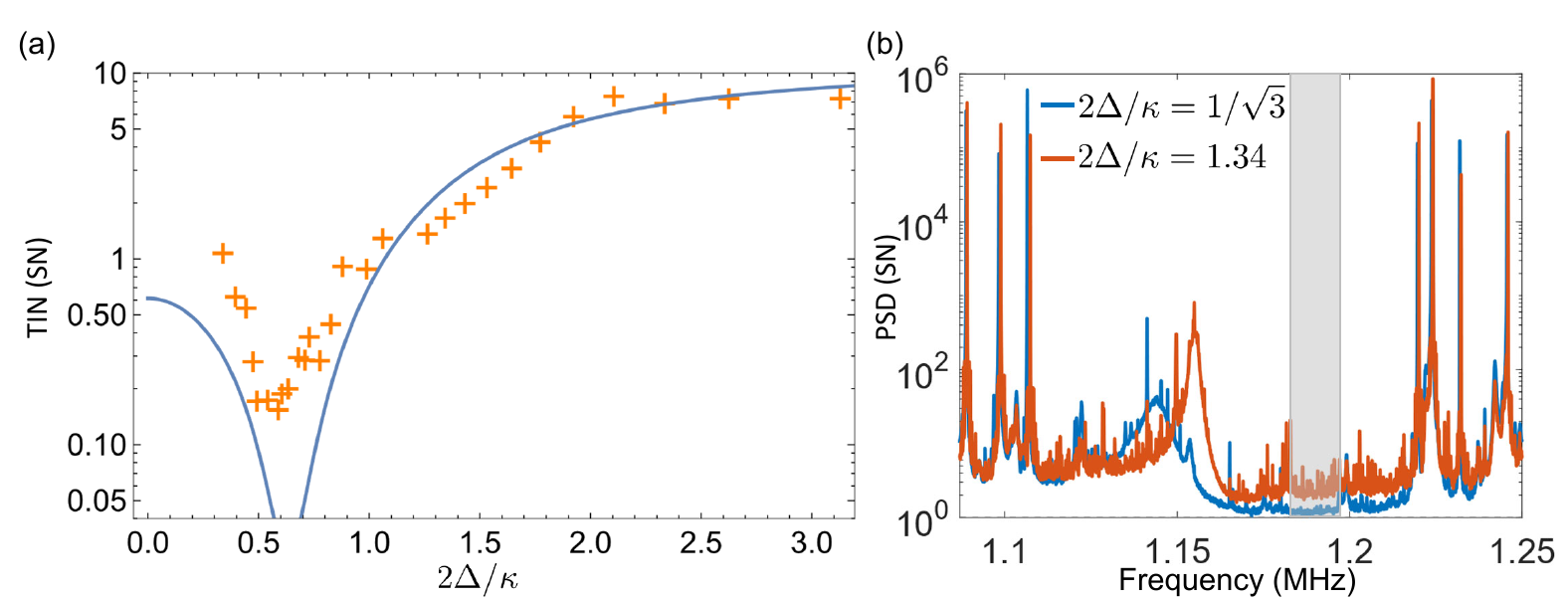}
    \caption{(a) TIN intracavity photon fluctuation measured with direct detection of the pump light at different cavity detunings. TIN is estimated by averaging the shaded  frequency band in (b).  (b) PSD of amplitude quadrature in shot noise (SN) unit at two cavity detunings $2\Delta/\kappa = 1/\sqrt{3}$ and $2\Delta/\kappa=1.34$. The noise floor at the magic detuning is slightly above the shot noise due to the  residual TIN.}
    \label{fig:S1}
\end{figure*}
In general, the dependence of TIN on the mean intracavity power $|\bar{a}|^2$ and cavity linewidth is complicated. On the one hand, more intracavity power (narrower cavity linewidth) helps cool all mechanical modes thus reducing $S_{\Delta\Delta}(\omega)$. On the other hand, it also increases the susceptibility of TIN (denominator in Eq. (S3)). We found the ratio of residual intensity noise of TIN to shot noise is roughly constant at magic detuning when varying the cooling power.

Fig.~\ref{fig:S1}. (a) shows the measured TIN in shot noise (SN) unit as a function of cavity detuning while the input power is fixed. The curve is fitted from the second order TIN model ( Eq.(S15)), where we assume a flat spectrum of $S_{\Delta\Delta}$ up to $7$ MHz. The TIN is estimated from the frequency band shown in the box in Fig.~\ref{fig:S1} (b).  The PSDs shown in Fig.~\ref{fig:S1} (b) are measured with optical direct detection using the cooling laser at 862nm. At magic detuning, we observe a broadband noise floor above the shot noise due to third-order TIN. We also see many mixing peaks away from the magic detuning that arise from the second-order TIN.

\subsection{Sideband cooling in the presence of residual thermal intermodulation noise}
\label{sec:network_theory}
Since TIN can not be canceled exactly at magic detuning in the limit of the slow cavity, its intensity fluctuations can drive the mechanical oscillator and raise its temperature. Here, we derive the effects of residual TIN in sideband cooling. In the Heisenberg picture, the quantum Langevin equations for the cavity mode $a$ and the dimensionless mechanical quadrature $\{\hat{Q},\hat{P}\}$  in the frequency domain are:
\begin{align}
\hat{a}(\omega) &= \chi_{\mathrm{cav}} \left[\omega)(-i\sqrt{2}g\hat{Q}(\omega)+\sum_i\sqrt{\kappa_i}\hat{a}_{\mathrm{in},i}(\omega)\right]+\delta a_{\mathrm{TIN}}(\omega)\\
\hat{a}^\dagger(\omega) &=[\hat{a}(-\omega)]^* = \chi^*_\mathrm{cav}(-\omega)\left[i\sqrt{2}g\hat{Q}(\omega)+\sum_i\sqrt{\kappa_i}\hat{a}^\dagger_{\mathrm{in},i}(\omega)\right]+\delta a_{\mathrm{TIN}}^\dagger(\omega)\\
-i\omega \hat{Q}(\omega) &= \Omega_m \hat{P}(\omega) \\
-i\omega \hat{P}(\omega) & = -\Omega_m \hat{Q}(\omega) -\Gamma \hat{P}(\omega)+\sqrt{2\Gamma}\hat{P}_{\mathrm{in}}(\omega)-\sqrt{2}g(\hat{a}(\omega)+\hat{a}^\dagger(\omega)) \\
\hat{Q}(\omega) &= \chi_{\mathrm{m}}(\omega) (\sqrt{2\Gamma}\hat{P}_{\mathrm{in}}(\omega)-2g X(\omega))= \hat{Q}_{\mathrm{th}}(\omega)-\chi_{\mathrm{m}}(\omega) 2g \hat{X}(\omega)
\end{align}
where we define cavity and mechanical susceptibility:
\begin{align}
    \chi_{\mathrm{cav}}^{-1}(\omega) &= \kappa/2-i(\Delta+\omega)\\
    \chi_\mathrm{m}(\omega) &= \frac{\Omega_m}{\Omega_m^2-\omega^2-i\omega\Gamma}
\end{align}
with laser detuning $\Delta = \omega_L-\omega_c$. $\hat{a}_{\mathrm{in},i}$ represents the input mode operators. We also include TIN as an additional cavity noise $\delta a_{\mathrm{TIN}}$. The dimensionless input momentum fluctuation  is characterized by the following spectral correlations: $S_{\hat{P}_{\mathrm{in}}\hat{P}_{\mathrm{in}}}(\omega) = \bar{n}_{\mathrm{th}}+1$ and $S_{\hat{P}_{\mathrm{in}}\hat{P}_{\mathrm{in}}}(-\omega) = \bar{n}_{\mathrm{th}}$ with $n_{\mathrm{th}}=k_B T/
\hbar\Omega_m$.
We hereafter remove the $\omega$ in the notation for simplicity. 

The dimensionless mechanical position quadrature is:
\begin{align}
    \hat{Q} = M[ \hat{Q}_{\mathrm{th}}-\sqrt{2}g\chi_\mathrm{m}(\sum_i\sqrt{\kappa_i}(\chi_{\mathrm{cav}} \hat{a}_{\mathrm{in},i}+\chi_{\mathrm{cav}}^*(-\omega) \hat{a}_{\mathrm{in},i}^\dagger)+\sqrt{2}X_{\mathrm{TIN}})]
\end{align}
and we define the following susceptibilities for simplicity:
\begin{align}
    \tilde{\chi}_c(\omega) = \chi_{\mathrm{cav}}^*(-\omega)-\chi_{\mathrm{cav}}(\omega)\\
    M(\omega) = (1+i\tilde{\chi}_c\chi_{\mathrm{m}}2g^2)^{-1}\\
\end{align}
The phonon occupation is thus
\begin{align}
    \bar{S}_{\hat{Q}\hat{Q}} = |\chi_\mathrm{m}'|^2 2\Gamma [n_{\mathrm{th}}+1/2+\frac{\kappa g^2}{2\Gamma}(|\chi_{\mathrm{cav}}(\omega)|^2+|\chi_{\mathrm{cav}}(-\omega)|^2)+\frac{2g^2}{\Gamma}\bar{S}_{X_{\mathrm{TIN}}X_{\mathrm{TIN}}}]
\end{align}

\noindent where $\chi_\mathrm{m}'=M\chi_\mathrm{m} \approx \Omega_m/(\Omega_m^2+2\omega\delta\Omega_{\mathrm{opt}}(\omega)-\omega^2-i\omega(\Gamma+\Gamma_{\mathrm{opt}}(\omega)))$ is the modified mechanical susceptibility with optomechanical damping rate $\Gamma_{\mathrm{opt}}$  and optical spring shift $\delta\Omega_{\mathrm{opt}}(\omega)$. We may also include an additional power-dependent mechanical frequency shift $\delta \Omega_{\mathrm{th}}\propto g^2$  and temperature change $T_{\mathrm{abs}}\propto g^2$ due to the absorption of  cooling beam. 

We also use the amplitude quadrature of cavity noise $X_{\mathrm{TIN}} = (\delta a_{\mathrm{TIN}}+\delta a^\dagger_{\mathrm{TIN}})/\sqrt{2}$. 
We choose $i=1$  as the transmission port of the cavity with coupling rate $\kappa_1\sim \kappa \gg \kappa_{i\neq1}$. From the input-output relations, the output field can be written as:
\begin{align}
    \hat{a}_{\mathrm{out}} &= \hat{a}_{\mathrm{in},1}-\sqrt{\kappa_1}\hat{a}\\
    &=i\sqrt{2}g\chi_{\mathrm{cav}} \sqrt{\kappa_1} \hat{Q} +(1-\kappa_1\chi_{\mathrm{cav}}) \hat{a}_{\mathrm{in},1}-\sum_{i\neq1}\sqrt{\kappa_1\kappa_i}\chi_{\mathrm{cav}}\hat{a}_{\mathrm{in},i}-\sqrt{\kappa_1}\delta a_{\mathrm{TIN}}
\end{align}
 When detecting the output field of the cooling beam, the interference between $\hat{Q}$ and $\delta a_{\mathrm{TIN}}$ will give fake sideband asymmetry. It should also be noted that spectral fitting will result in overestimated $g$  without including this interference term. The explicit expression is given by:
\begin{align}
   \hat{a}_{\mathrm{out}} &= i\sqrt{2}g\chi_{\mathrm{cav}} \sqrt{\kappa_1} M[\hat{Q}_{\mathrm{th}}-\sqrt{2}g\chi_\mathrm{m}(\sum_i\sqrt{\kappa_i}(\chi_{\mathrm{cav}} \hat{a}_{\mathrm{in},i}+\chi_{\mathrm{cav}}^*(-\omega)\hat{a}_{\mathrm{in},i}^\dagger))] \\ & -\sqrt{\kappa_1}\delta a_{\mathrm{TIN}}-i2\sqrt{2}g^2\chi_{\mathrm{cav}}\sqrt{\kappa_1}M\chi_\mathrm{m} X_{\mathrm{TIN}} +\frac{-i(\Delta+\omega)-(\kappa_1-\kappa/2)}{-i(\Delta+\omega)+\kappa/2} \hat{a}_{\mathrm{in},1}-\sum_{i\neq1}\sqrt{\kappa_1\kappa_i}\chi_{\mathrm{cav}}\hat{a}_{\mathrm{in},i}
\end{align}

\subsection{Sideband asymmetry from a dual homodyne measurement with a weak probe}
\label{sec:benchmark}
To avoid the TIN of the cooling beam at the detector, we use another weak probe laser ($C_q \ll 1$) at 813 nm with a cavity linewidth $49.3$ MHz. This weak probe allows us to measure different optical quadratures at the shot noise limit via a single-port homodyne detection where TIN noise can be cancelled by tuning the LO amplitude and phase\cite{Huang2024,fedorov2020thermal}. To reconstruct the sideband asymmetry, we implement a dual-homodyne measurement instead of a conventional heterodyne detection. We split the probe light evenly into two optical paths and measure two optical quadratures at homodyne angles $\theta_1$ and $\theta_2$ simultaneously. Specifically, we measured the optical amplitude quadrature $\theta_1=0$ with direct detection at one detector. The TIN at amplitude quadrature is cancelled at the magic detuning. To measure the second quadrature $\theta_2\approx -\pi/3$, we implement a single port homodyne detection $a_d \approx a_{\mathrm{sig}}+r a_{\mathrm{LO}}$ which combines the signal field $a_{sig}$ and local oscillator field $a_{\mathrm{LO}}$  at a highly asymmetrical beam splitter with reflectivity $r\ll1$. The TIN at phase quadrature $\theta_2$ is cancelled by tuning the LO oscillator amplitude while locking to phase quadrature $\theta_2$.  The field operator $ \hat{a} = \hat{X}+i\hat{Y}$  can be reconstructed from the two homodyne measurement records $\hat{X}_{\theta_1},\hat{X}_{\theta_2}$:
\begin{align}
    \hat{X}_{\theta_1} = \sqrt{\eta_1}(\cos{\theta_1}\hat{X}+\sin{\theta_1}\hat{Y})+\sqrt{1-\eta_1}\hat{X}_{vac,1}\\
    \hat{X}_{\theta_2} = \sqrt{\eta_2}(\cos{\theta_2}\hat{X}+\sin{\theta_2}\hat{Y})+\sqrt{1-\eta_2}\hat{X}_{vac,2}
\end{align}
This extra freedom of LO amplitude for different optical quadratures allows us to measure the sideband asymmetry without TIN contamination in contrast to a heterodyne detection.

Now we include the weak probe laser $\hat{a}_{\mathrm{p}}$ in the coupled quantum Langevin equations of motion:

\begin{align}
    &\hat{a}_{\mathrm{p}}= \chi_{\mathrm{p}}(\omega) (-i\sqrt{2}g_\mathrm{p} \hat{Q}+\sum_j\sqrt{\kappa_{\mathrm{p},j}}\hat{a}_{\mathrm{p,in},j} )\\
    &\hat{a}_{\mathrm{p,out}} = \hat{a}_{\mathrm{p,in},1}-\sqrt{\kappa_{\mathrm{p},1}}b = \chi_{\mathrm{p}}(\omega) i\sqrt{2}g_{\mathrm{p}}\sqrt{\kappa_{\mathrm{p},1}} \hat{Q} +(1-\kappa_{\mathrm{p},1}\chi_{\mathrm{p}}) \hat{a}_{\mathrm{p,in},1}-\sum_{j\neq 1}\sqrt{\kappa_{\mathrm{p},1}\kappa_{\mathrm{p},j}}\chi_{\mathrm{p}} \hat{a}_{\mathrm{p,in},j}\\
    &\hat{Q} = \chi_\mathrm{m}'[\sqrt{2\Gamma}\hat{P}_{\mathrm{in}}-\sum_i\sqrt{2\kappa_i}g(\chi_{\mathrm{cav}}\hat{a}_{\mathrm{in},i}+\chi_{\mathrm{cav}}^*(-\omega)\hat{a}^\dagger_{\mathrm{in},i})-\sum_j\sqrt{2\kappa_{\mathrm{p},j}}g_{\mathrm{p}}(\chi_{\mathrm{p}}\hat{a}_{\mathrm{p,in},j}+\chi_{\mathrm{p}}^*(-\omega)\hat{a}^\dagger_{\mathrm{p,in},j})+2gX_{\mathrm{TIN}}]
\end{align}
$\chi_{\mathrm{p}}^{-1} = \kappa_{\mathrm{p}}/2-i(\Delta_{\mathrm{p}}+\omega) $ is the cavity susceptibility of mode $\hat{a}_{\mathrm{p}}$.  And the new mechanical susceptibility including the backaction of two beams is :
\begin{align}
   \chi_\mathrm{m}'^{-1}= (\chi_\mathrm{m}^{-1}+2ig_{\mathrm{p}}^2(\chi_{\mathrm{p}}^*(-\omega)-\chi_{\mathrm{p}}(\omega))+2ig^2(\chi_{\mathrm{cav}}^*(-\omega)-\chi_{\mathrm{cav}}(\omega)))
\end{align}
The symmetrized PSD of $\hat{a}_{\mathrm{p,out}}$ is:
\begin{align}
    \bar{S}_{\hat{a}_{\mathrm{p,out}}\hat{a}_{\mathrm{p,out}}} (\omega)  &= \{a_{\mathrm{p,out}}^{\dagger}(\omega),\hat{a}_{\mathrm{p,out}}(-\omega)\}/2\\
    &= 2\eta g_{\mathrm{p}}^2\kappa_{\mathrm{p},1}|\chi_{\mathrm{p}}(-\omega)|^2\bar{S}_{\hat{Q}\hat{Q}}+2\eta g_{\mathrm{p}}^2\kappa_{\mathrm{p},1} |\chi_{\mathrm{p}}(-\omega)|^2 Im\{\chi_\mathrm{m}' \}+1/2
\end{align}
where the middle term comes from the quantum correlation between the backaction and imprecision. We also include measurement efficiency as $\eta$. By virtue of the equipartition theorem, we can write the sideband asymmetry in terms of the phonon number:
\begin{align}
    R = \frac{\int_0^\infty (\bar{S}_{\hat{a}_{\mathrm{p,out}}\hat{a}_{\mathrm{p,out}}} (\omega) -1/2)d\omega}{\int_{-\infty}^0 (\bar{S}_{\hat{a}_{\mathrm{p,out}}\hat{a}_{\mathrm{p,out}}} (\omega) -1/2)d\omega}  \approx\frac{|\chi_{\mathrm{p}}(-\Omega'_m)|^2 (\bar{n}+1)}{|\chi_{\mathrm{p}}(\Omega'_m)|^2 \bar{n}}
\end{align}

where $\Omega'_m = \Omega_m+\delta\Omega_{\mathrm{opt}}$ is the shifted mechanical frequency.

\section{Experimental details}

\subsection{Power spectral density calibration}
\begin{figure*}
    \centering
    \includegraphics[width=1\textwidth]{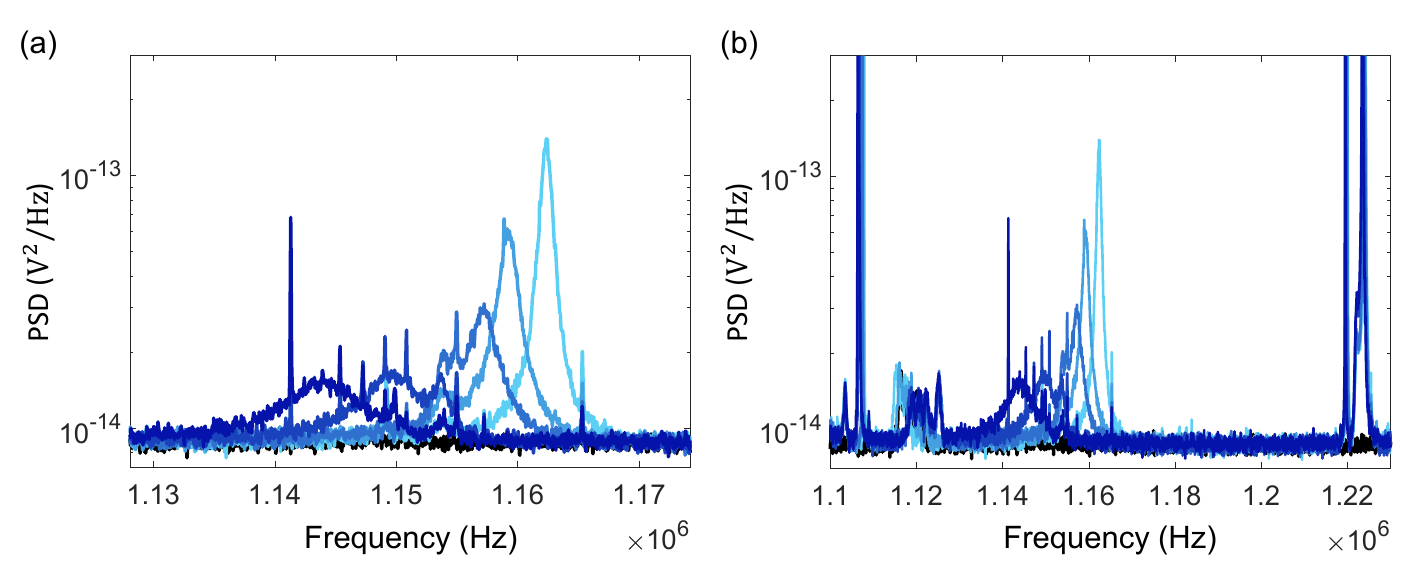}
    \caption{PSD of 813nm measured with direct detection at various cooling powers of 861nm, increasing from light to dark blue. (a) zoom in of defect mode in (b). Black: shot noise. }
    \label{fig:S2}
\end{figure*}

In the experiment, we use two Ti:Sapphire lasers to cool and probe the mechanical resonator. The laser intensity noise and phase noise are characterized elsewhere\cite{Huang2024} and their contribution to the shot noise of probe light is less than $1\%$. Hence we ignore them in the analysis. The PSDs of direct detection with probe light at different cooling powers are shown in Fig.~\ref{fig:S2} where the black line is the shot noise floor.  

\subsection{Calibration of dual homodyne detectors}

To calibrate the dual homodyne detection, we split the output light by 50/50 and measure both optical beams with direct detection. The photocurrents from both detectors are digitized simultaneously at a sampling rate of $31.25$ MHz and Fourier transformed to the frequency domain. We use mechanical peaks as calibration tones to fit the gain ratio and phase delay between the two detectors which will be used to rescale the photocurrents before combining them to derive the sideband asymmetry. 

\begin{figure*}
    \centering
    \includegraphics[width=1\textwidth]{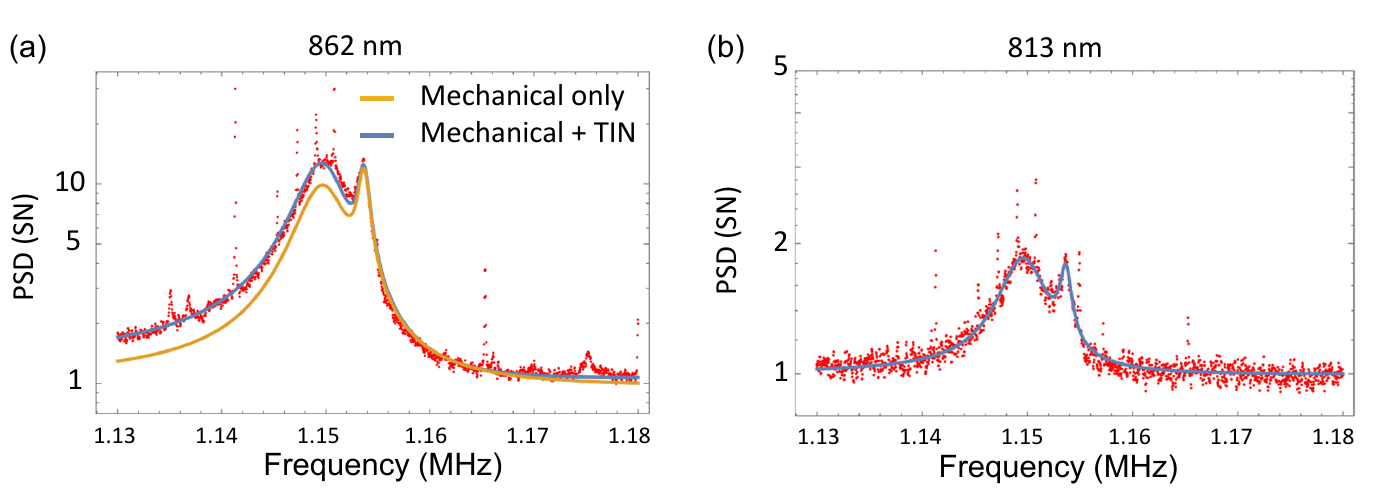}
    \caption{Power spectral density (normalized to shot noise) acquired with direct detection of cooling beam (a) and probe beam (b). The TIN from the cooling laser can drive the mechanical oscillator and interfere with itself at the detector, showing a broader spectral width (blue line) compared to purely optomechanical damping rate (orange line) fitted from the PSD of probe laser in (a).}
    \label{fig:S3}
\end{figure*}

\subsection{Backaction from the residual TIN of cooling beam}

Here we show that the TIN from the cooling beam can drive the mechanical resonator and interfere with itself at the detector. Fig.~\ref{fig:S3} shows the PSD normalized to shot noise acquired with direct detection of both cooling beam (a) and probe beam (b). We fit the optomechanical damping rate $5526$ Hz from the PSD of the probe light which is free from TIN. It is only possible to fit the PSD of the pump light with the contribution of TIN. We note that the spectral width is larger than the optomechanical damping rate fitted from the PSD of the probe light which suggests direct spectral fitting of the PSD of the pump laser overestimates the optomechanical damping rate.

\end{appendix}

\twocolumngrid

\bibliography{Sideband_Asymmetry}

\end{document}